\documentclass[showpacs,amssymb,aps]{revtex4}
\usepackage{footmisc}
\usepackage{amsmath}
\usepackage{amstext}
\usepackage{amsopn}
\usepackage{amsfonts}
\usepackage{amssymb}
\usepackage{bbm}
\usepackage{accents}
\usepackage{empheq}
\usepackage{graphicx}
\usepackage{epsf}
\usepackage{graphics}
\usepackage[latin1]{inputenc}

\renewcommand{\thefootnote}{\fnsymbol{footnote}}
\long\def\symbolfootnote[#1]#2{\begingroup%
\def\thefootnote{\fnsymbol{footnote}}\footnote[#1]{#2}\endgroup}
\newcommand{\be}{\begin{equation}}
\newcommand{\ee}{\end{equation}}
\newcommand{\bea}{\begin{eqnarray}}
\newcommand{\eea}{\end{eqnarray}}
\newcommand{\beaa}{\begin{eqnarray*}}
\newcommand{\eeaa}{\end{eqnarray*}}

\newcommand{\BB}{{{\rm I} \kern -2pt \rlap {\rm B} \kern +8pt}}

\begin{document}
\title{Darboux transformation and multi-soliton solutions of
Two-Boson hierarchy}

\author{Ashok Das $^{a,b}$ and U. Saleem $^{a,c}$\footnote{$\ $
e-mail: das@pas.rochester.edu, usaleem@physics.pu.edu.pk, usaleem@pas.rochester.edu}}

\affiliation{$^a$ Department of Physics and Astronomy, University of Rochester, Rochester, NY
14627-0171, USA}
\affiliation{$^b$ Saha Institute of Nuclear Physics, 1/AF Bidhannagar, Calcutta 700064, India}

\affiliation{$^c$ Department of Physics, University of the Punjab, Quaid-e-Azam
Campus, Lahore-54590,Pakistan}

\begin{abstract}
We study Darboux transformations for the two boson (TB) hierarchy both in the scalar as well as in the matrix descriptions of the linear equation. While Darboux transformations have been extensively studied for integrable models based on $SL(2,R)$ within the AKNS framework, this model is based on $SL(2,R)\otimes U(1)$. The connection between the scalar and the matrix descriptions in this case implies that the generic Darboux matrix for the TB hierarchy has a different structure from that in the models based on $SL(2,R)$ studied thus far. The conventional Darboux transformation is shown to be quite restricted in this model. We construct a modified Darboux transformation which has a much richer structure and which also allows for multi-soliton solutions to be written in terms of Wronskians. Using the modified Darboux transformations, we explicitly construct  one soliton/kink solutions for the model.
\end{abstract}
\bigskip

\pacs{02.30.lk, 05.45.Yv, 11.10.-z}


\maketitle

\bigskip

\section{Introduction}
During the past four decades there has been a lot of interest  in
the study of different classical and quantum integrable models
from various points of view, for example, the Lax description,
zero-curvature condition, the bi-Hamiltonian structures,
multi-soliton solutions, conserved quantities etc.
\cite{Ablowitz,Babelon}. Most of the familiar $1+1$ dimensional
integrable systems can be described (in the matrix form) by the
AKNS formalism, namely, a zero-curvature description for them is
based on the symmetry group $SL (2, R)$. The multi-soliton
solutions for such systems
 correspondingly have been constructed (among other methods) by the conventional Darboux
 transformations \cite{darboux, matveev} (in the scalar description) or in the matrix formalism (zero-curvature method) by Darboux transformations based on the symmetry group $SL (2, R)$\cite{Gu}.

The Two Boson (TB) hierarchy is an integrable system which in many
ways is quite different from the other familiar integrable systems
and has been studied extensively during the last two decades
\cite{TBsystem1}-\cite{roy}. For example, the scalar Lax
description of this model involves a nonstandard equation
\cite{TBsystem1} and this system is known to be tri-Hamiltonian.
In fact, the Hamiltonian structures of the bosonic as well as the
supersymmetric TB hierarchy have been studied in
\cite{TBsystem1,TBsystem11} and the bilinear and the trilinear
forms of the hierarchy have been used in
\cite{TBsystem0}-\cite{TBsystem12} to construct the multi-soliton
solutions of this system. Furthermore, the zero-curvature
formulation of this sytem (leading to the tri-Hamiltonian
structure) does not fall within the conventional AKNS hierarchy
based on $SL (2, R)$, rather it is based on the symmetry algebra
$SL (2, R)\otimes U(1)$ \cite{roy}. As a result, one expects the
Darboux transformation for this system to have new features and in
this letter we show that this expectation is indeed true. In
particular, we show that the conventional Darboux transformation
in this case is quite restricted while a modified  Darboux
transformation leads to a much richer structure for solutions.
Furthermore, the relation between the scalar and the matrix
descriptions of the linear equations in this case leads to a
Darboux matrix which is generically quite distinct from that
studied so far within the context of the AKNS formalism. In
section {\bf II}, we recapitulate briefly various properties of
interest of the TB hierarchy. In section {\bf III}, we construct
the conventional $N$-fold Darboux transformation (both for the
scalar Lax equation and the zero-curvature condition) and express
the multi-soliton solutions for this system in terms of
Wronskians. In section {\bf IV},  we introduce a modified $N$-fold
Darboux transformation which has a much richer structure which
also allows the multi-soliton solutions to be written in terms of
the corresponding Wronskians. In the last section, we derive
explicitly the one soliton/kink solutions of the TB hierarchy
using the modified Darboux transformation.

\section{Recapitulation of the TB hierarchy}

The TB equation is a set of two $1+1$ dimensional equations given by \cite{TBsystem1,TBsystem11}
\begin{eqnarray}
J_{0t}&=&\left(2J_{1}+J_{0}^{2}-J_{0x}\right)_{x},\nonumber\label{TB1}\\
J_{1t}&=&\left(2J_{0}J_{1}+J_{1x}\right)_{x},\label{TB2}
\end{eqnarray}
where $J_{0}, J_{1}$ represent the dynamical variables (depending on $t$ and $x$) of the system and the subscripts $t$ and $x$ represent partial derivatives with respect to these variables.
The TB system (\ref{TB2}) can be described as a nonstandard scalar Lax equation of the form
\begin{eqnarray}
\frac{\partial L}{\partial t}&=&\left[L,M\right],\label{Lax1}
\end{eqnarray}
where the Lax operators $L$ and $M$ are given by
\begin{eqnarray}
L=\partial-J_{0}+\partial^{-1}J_{1},\quad\quad\quad M\equiv
\left(L^{2}\right)_{\geq1}=-\partial^{2}+2J_{0}\partial.\label{Lax2}
\end{eqnarray}
Here $\partial=\frac{\partial}{\partial x}$ and $\partial^{-1}$
denotes the formal integration operation. The Lax operators $L$ and $M$
lead to the linear equations
\begin{eqnarray}
L\psi=\lambda\psi,\quad\quad\quad \psi_{t}=M\psi,\label{Lax3}
\end{eqnarray}
where $\lambda$ is the constant spectral parameter and
$\psi=\psi\left(x,t;\lambda\right)$ is a scalar wave function. Equation \eqref{Lax3} generates the TB hierarchy of equations.

The linear equations (\ref{Lax3}) have the explicit forms
\begin{eqnarray}
\psi_{x}=\left(J_{0}+\lambda\right)\psi-\partial^{-1}\left(J_{1}\psi\right),\quad\quad\quad
\psi_{t}=\left(J_{0}-\lambda\right)\psi_{x}+\left(J_{1}-J_{0x}\right)\psi,\label{linear2}
\end{eqnarray}
where the second equation has been simplified using the first and the compatibility condition (i.e. $\psi_{xt}=\psi_{tx}$) of the linear system (\ref{linear2}) is equivalent to the Lax equation \eqref{Lax1} and leads to the TB system (\ref{TB2}). For future use, we note that we can avoid the nonlocality in the first equation in (\ref{linear2}) by operating with a derivative and, therefore, the pair of linear equations can equivalently be written as
\begin{eqnarray}
\psi_{xx}=\left(J_{0}+\lambda\right)\psi_{x}+\left(J_{0x}-J_{1}\right)\psi,\quad\quad\quad
\psi_{t}=\left(J_{0}-\lambda\right)\psi_{x}+\left(J_{1}-J_{0x}\right)\psi.\label{linear4}
\end{eqnarray}
It can be checked that the compatibility condition for \eqref{linear4} (i.e. $\psi_{xxt}=\psi_{txx}$)
leads to the TB system of equations (\ref{TB2}).

The TB hierarchy of equations \eqref{linear2} can also be expressed as a zero-curvature
condition in terms of Lie algebra valued gauge fields $A_{0}$ and $A_{1}$ based on $SL(2,R)\otimes U(1)$ \cite{roy},
\begin{eqnarray}
\left[\partial_{x}-A_{1},\partial_{t}-A_{0}\right]\equiv
\partial_{t} A_{1}-\partial_{x} A_{0}-\left[A_{0}, A_{1}\right]=0,\label{zero5}
\end{eqnarray}
where
\begin{equation}
A_{1}(x,t;\lambda) = \left(\begin{array}{cc}
J_{0}+\lambda & -1 \\
J_{1} & 0%
\end{array}\right),\quad
A_{0}(x,t;\lambda) = \left(\begin{array}{cc}
J^{2}_{0}+J_{1}-J_{0x}-\lambda^{2} & \lambda-J_{0} \\
J_{0}J_{1}+J_{1x}-\lambda J_{1} & J_{1}%
\end{array}\right).\label{zero4}
\end{equation}
The zero-curvature condition (\ref{zero5}) can be understood as the
compatibility condition for the linear (matrix) equations
\begin{eqnarray}
\partial_{x} \Psi(x,t;\lambda)=A_{1}(x,t;\lambda)\Psi(x,t;\lambda),\quad\quad\quad
\partial_{t} \Psi(x,t;\lambda)=A_{0}(x,t;\lambda)\Psi(x,t;\lambda),\label{zero2}
\end{eqnarray}
where $\Psi(x,t;\lambda)$ denotes a $SL(2,R)\otimes U(1)$ Lie algebra valued  matrix wave function. The connection between scalar linear equations (of the form \eqref{linear2}) and matrix equations (of the form \eqref{zero2}) can be traced to the work of Drinfeld and Sokolov \cite{drinfeld,daslecture}. We note here for future use that the linear equations \eqref{linear4} can be related to the matrix equations (following \cite{drinfeld,daslecture})
\begin{eqnarray}
\partial_{x}\Psi(x,t;\lambda)=B_{1}(x,t;\lambda)\Psi(x,t;\lambda),\quad\quad\quad
\partial_{t}\Psi(x,t;\lambda)=B_{0}(x,t;\lambda)\Psi(x,t;\lambda),\label{zero7}
\end{eqnarray}
where the $SL (2, R)\otimes U(1)$ valued gauge fields $B_{0}, B_{1}$ have the forms
\begin{eqnarray}
B_{1}(x,t;\lambda)=\left(\begin{array}{cc}
J_{0}+\lambda & J_{0x}-J_{1} \\
1 & 0%
\end{array}\right),\quad
B_{0}(x,t;\lambda)=\left(\begin{array}{ccc}
J^{2}_{0}+J_{1}-\lambda^{2} &  &\left(\lambda-J_{0}\right)\left(J_{1}-J_{0x}\right)+\left(J_{1}-J_{0x}\right)_{x} \\
J_{0}-\lambda  &  & J_{1}-J_{0x}%
\end{array}\right).\label{zero9}
\end{eqnarray}
The compatibility condition of the matrix linear system
(\ref{zero7}) corresponds to the zero-curvature condition
\begin{eqnarray}
\left[\partial_{x}-B_{1},\partial_{t}-B_{0}\right]\equiv
\partial_{t} B_{1}-\partial_{x}B_{0}-\left[B_{0}, B_{1}\right]=0.\label{zero10}
\end{eqnarray}

\section{Conventional Darboux transformation}

In this section, we study systematically the properties of the conventional Darboux transformations for the scalar wave function $\psi$ as well as the matrix wave function $\Psi$ of the linear systems \eqref{linear4} and \eqref{zero2} (or equivalently \eqref{zero7}) associated with the TB hierarchy.

Let us recall that Darboux transformation allows one to construct a new solution of a Sturm-Liouville problem from a known solution. Alternatively, it can be used to obtain the solution of a Sturm-Liouville system from another (Sturm-Liouville) system. Conventionally, the Darboux transformation is defined as a transformation of the (scalar) wave function $\psi(x,t;\lambda )$ of the Sturm-Liouville system of the form
\begin{eqnarray}
&& \psi(x,t;\lambda )\longmapsto
\psi'(x,t;\lambda,\lambda_{1})=\psi_{x}-\sigma_{1}\psi,\label{Darboux1}
\end{eqnarray}%
where
\begin{equation}
\sigma_{1}=\frac{\psi_{1x}}{\psi_{1}},\label{sigma1}
\end{equation}
and $\psi_{1}$ is a solution of the original system for
$\lambda=\lambda_1$. The new solution $\psi'(x,t;\lambda,\lambda_{1})$ clearly satisfies the condition
\begin{eqnarray}
\psi'(x,t;\lambda,\lambda_{1})|_{\lambda=\lambda_{1}}=0.\label{Darboux2}
\end{eqnarray}%
Requirement of covariance of the linear equation under the Darboux transformation \eqref{Darboux1} then determines the transformations of the potentials (other variables) in the problem. While it is not obvious in the form \eqref{Darboux1}, Darboux transformation can be thought of as a very special gauge transformation which becomes manifest only in the context of the matrix description of a linear equation.

The conventional Darboux transformation \eqref{Darboux1} has been generalized to the AKNS family of integrable models (based on $SL(2,R)$) and is one of the well-known methods of obtaining multi-soliton solutions for such systems (see, for example, \cite{darboux}-\cite{Gu}). Let us now apply the conventional Darboux transformation \eqref{Darboux1} to the linear equations \eqref{linear4}. Requirement of covariance of the linear equation \eqref{linear4} under \eqref{Darboux1} leads to the fact that $\psi'(x,t;\lambda,\lambda_{1})$ satisfies the linear equations
\begin{eqnarray}
\psi'_{xx}=\left(J'_{0}+\lambda\right)\psi'_{x}+\left(J'_{0x}-J'_{1}\right)\psi',\quad\quad\quad
\psi'_{t}=\left(J'_{0}-\lambda\right)\psi'_{x}+\left(J'_{1}-J'_{0x}\right)\psi',\label{DTlinear2}
\end{eqnarray}
which in turn determines the transformations of the potentials (under the Darboux transformation) to be
\begin{eqnarray}
J'_{0}= J_{0},\quad\quad\quad J'_{1}=J_{1}-J_{0x}+2\sigma_{1x} =
J_{1}-J_{0x}+2\partial^{2}\ln \psi_{1}.\label{Darboux3}
\end{eqnarray}
It can be checked that the transformed variables in \eqref{Darboux3} do satisfy the TB equation (see \eqref{TB1})
\begin{eqnarray}
J'_{0t}=\left(2J'_{1}+(J'_{0})^{2}-J'_{0x}\right)_{x},\quad\quad
J'_{1t}=\left(2J'_{0}J'_{1}+J'_{1x}\right)_{x},\label{DTTB2}
\end{eqnarray}
which also arises from the compatibility of \eqref{DTlinear2}. This shows that with some starting (seed) solution $(J_{0},J_{1})$ of the two boson equation, one can construct new solutions $(J'_{0},J'_{1})$ through the Darboux transformation. However, it is interesting to note from \eqref{Darboux3} that under the conventional Darboux transformation (\ref{Darboux1}) the dynamical variable $J_{0}$ remains invariant so that the conventional Darboux transformation in this case is quite restrictive. Therefore, to summarize, we note that if the set $(\psi,J_{0},J_{1})$ defines a solution of the linear equations (\ref{linear4}),
then the set $(\psi[1],J_{0}[1],J_{1}[1])$ defined by
\begin{eqnarray}
\psi[1]&=&\psi' \equiv \psi_{x}-\frac{\psi_{1x}}{\psi_{1}}\psi=\frac{W(\psi_{1},\psi)}{\psi_{1}},\label{DTfield}\\
J_{0}[1]&=&J_{0},\label{DTfielda}\\
J_{1}[1]&=&J_{1}-J_{0x}+2\sigma_{1x} = J_{1}-J_{0x}+2\partial^{2}\ln \psi_{1},
\label{DTfieldb}
\end{eqnarray}%
is the required one-fold Darboux transformation of the set under
which \eqref{linear4} is covariant. Here
\begin{equation}
W(\psi_{1},\psi)\equiv\left|\begin{array}{cc}
\psi_{1} & \psi \\
\partial \psi_{1} & \partial \psi%
\end{array}\right|=\psi_{1}\psi_{x}- \psi_{1x} \psi,\label{wronskian1}
\end{equation}
is the usual Wronskian determinant of the two solutions.

The two-fold Darboux transformation is constructed in two steps.
First, with a solution $\psi_{1}$ of the linear system
(\ref{linear4}) for the eigenvalue $\lambda_{1}$ we construct the
set $(\psi[1],J_{0}[1],J_{1}[1])$ as discussed in
\eqref{DTfield}-\eqref{DTfieldb}. Then, with a solution
$\psi_{2}[1]$ of \eqref{DTlinear2} for eigenvalue $\lambda_{2}$,
the two-fold Darboux transformation on the set $(\psi,
J_{0},J_{1})$ is determined to be
\begin{eqnarray}
\psi[2]&\equiv&\left(\psi[1]\right)_{x}-\frac{\left(\psi_{2}[1]\right)_{x}}{\psi_{2}[1]}\psi[1]
=\frac{W(\psi_{1},\psi_{2},\psi)}{W(\psi_{1},\psi_{2})},\nonumber\\
J_{0}[2]&\equiv& J_{0}[1]=J_{0}, \nonumber\\
J_{1}[2]&\equiv&J_{1}[1]-J_{0x}[1]+2\partial^{2}\ln
\psi_{2}[1]=J_{1}-2J_{0}+2\partial^{2}\ln W(\psi_{1},\psi_{2}),
\label{twoDTfieldb}
\end{eqnarray}
where
\begin{equation}
W(\psi_{1},\psi_{2},\psi)\equiv\left|\begin{array}{ccc}
\psi_{1}&\psi_{2} & \psi \\
\partial \psi_{1} & \partial \psi_{2} & \partial \psi\\
\partial^{2} \psi_{1} & \partial^{2} \psi_{2} & \partial^{2} \psi
\end{array}\right|, \label{wronskian2}
\end{equation}
and $W(\psi_{1},\psi_{2})$ is defined in \eqref{wronskian1}. This procedure can be iterated $N$ times leading to the $N$-fold
Darboux transformation on the set $(\psi,J_{0},J_{1})$ in the form
\begin{eqnarray}
\psi[N] &=& \frac{W(\psi_1,\psi_2,\dots,\psi_N,\psi)}{W(\psi_1,\psi_2,\dots,\psi_N)}, \nonumber\\
J_{0}[N]&=&J_{0},\nonumber\\
J_{1}[N]&=&J_{1}-NJ_{0x}+2\partial^{2}\ln
W(\psi_1,\psi_2,\dots,\psi_N),\label{NDTfieldb}
\end{eqnarray}
where
\begin{equation}
W(\psi_1,\psi_2,\dots,\psi_N,\psi) = \left\vert
\begin{array}{cccccc}
\psi_{1}& \psi_{2}  & \cdots  & & \psi_{N}& \psi \\
\partial\psi_{1}& \partial\psi_{2}  & \cdots  & & \partial\psi_{N}& \partial\psi\\
\partial^{2}\psi_{1}& \partial^{2}\psi_{2}  & \cdots  & & \partial^{2}\psi_{N}& \partial^{2}\psi \\
\vdots  & \vdots  & \ddots  & \vdots  & \vdots & \vdots \\
\partial^{N}\psi_{1}& \partial^{N}\psi_{2}  & \cdots  & & \partial^{N}\psi_{N}& \partial^{N}\psi
\end{array}%
\right\vert.\label{wronskian3}
\end{equation}
It is worth noting here that under a conventional Darboux
transformation  (\ref{Darboux1}) one of the dynamical variables,
$J_{1}$, transforms while the second dynamical variable, $J_{0}$,
remains invariant. If we set $J_{0}=0$ and $J_{1}=u$, the TB
hierarchy \eqref{linear4} is known to reduce to the KdV hierarchy
and in this case the $N$-fold Darboux transformation
(\ref{NDTfieldb}) of the dynamical variable $J_{1}$ indeed reduces
to the $N$-fold Darboux transformation of the KdV variable $u$.

The Darboux transformations for the matrix solution $\Psi$ of the
linear system (\ref{zero2}) or \eqref{zero7} are given in terms of
a $2\times 2$ matrix $D(x,t;\lambda,\lambda_{1} )$, called the
Darboux matrix. For a general discussion on Darboux matrix
approach see e.g. \cite{Gu}. However, since we have already
determined the (conventional) Darboux transformation for the
scalar linear equation \eqref{linear4} (see
\eqref{DTfield}-\eqref{DTfieldb}), the Darboux matrix can be
simply constructed following the connection between the scalar and
the matrix descriptions for this system \cite{daslecture} and this
brings out some new features. For example, if we consider the
linear matrix equation \eqref{zero2}, the one-fold Darboux
transformation of the matrix wave function has the form
\begin{equation}
\Psi'(x,t;\lambda,\lambda_{1})= D(x,t;\lambda,\lambda_{1}
)\Psi(x,t;\lambda ), \label{MatrixDarboux1}
\end{equation}%
and the Darboux matrix has the form
\begin{equation}
D (x,t;\lambda,\lambda_{1}) = \begin{pmatrix}
J_{0}+\lambda-\sigma_{1} & -1\\
-J_{0x} + J_{1}+\sigma_{1x} & -\sigma_{1}
\end{pmatrix},
\end{equation}
where $\sigma_{1}$ is defined in \eqref{sigma1}. Furthermore, using \eqref{zero4} and \eqref{zero2}, we note that the Darboux transformation \eqref{MatrixDarboux1} for the wave function can be equivalently written as (which is simpler for calculations)
\begin{equation}
\Psi'(x,t;\lambda,\lambda_{1})\equiv D(x,t;\lambda,\lambda_{1}
)\Psi(x,t;\lambda
)=\left(\partial-M(x,t;\lambda_{1})\right)\Psi(x,t;\lambda ),
\label{MatrixDarboux1a}
\end{equation}
where the matrix $M(x,t;\lambda_{1})$ is given by (here one uses
the equation satisfied by $\sigma_{1}$ to simplify the matrix into
the following form)
\begin{equation}
M(x,t;\lambda_{1})=\left(\begin{array}{cc} \sigma_{1} &
0\\
\sigma_{1}^2-(J_{0}+\lambda_1)\sigma_{1}+J_{1} & \sigma_{1}%
\end{array}\right).\ \ \label{Darboux2a}
\end{equation}
It is worth pointing out here that the Darboux matrix in the AKNS framework (based on $SL(2,R)$) has the generic form
\begin{equation}
D = \lambda\mathbbm{1} - M\label{generic},
\end{equation}
where $\mathbbm{1}$ denotes the $2\times 2$ identity matrix. However, it can be checked that a Darboux matrix such as in \eqref{generic} does not work in the case of the TB hierarchy (and, in fact, can be simply understood from the connection between the scalar and the matrix descriptions). On the other hand, the Darboux matrix for the TB hierarchy has the generic form as given in \eqref{MatrixDarboux1a}.

The requirement of covariance of \eqref{zero2}, requires that the transformed matrix wave function $\Psi'(x,t;\lambda,\lambda_{1})$ satisfies
\begin{eqnarray}
\frac{\partial \Psi'}{\partial x}   =A_{1}' \Psi' ,\quad \quad
\quad \frac{\partial \Psi'}{\partial t}= A_{0}'\Psi'
,\label{MatrixDarboux2}
\end{eqnarray}%
where the transformation of the gauge potentials (fields) $A_{1}'$ and $A_{0}'$ are given by
\begin{eqnarray}
A_{1}'= D A_{1} D^{-1}+D_{x}D^{-1},\quad \quad \quad A_{0}'= D
A_{0} D^{-1}+D_{t}D^{-1}.\label{MatrixDarboux3}
\end{eqnarray}
The zero-curvature condition \eqref{zero5} is covariant under a
gauge transformation so that the transformed potentials
$(A_{1}',A_{0}')$ lead to a vanishing field strength (curvature).
Under the Darboux transformation \eqref{MatrixDarboux1a} and
\eqref{MatrixDarboux3}, the set $(\Psi,A_{1},A_{0})$ maps into
$(\Psi', A'_{1}, A'_{0})$.

Similarly, the matrix Darboux transformation for the linear matrix equation (\ref{zero7}) can be written in the generic form (\ref{MatrixDarboux1a})  with
\begin{equation}
M(x,t;\lambda_{1})= \left(\begin{array}{cc} \sigma_{1} &
-\sigma_{1}^2+(J_0+\lambda_{1})\sigma_{1}+(J_{0x}-J_{1})\\
0 & \sigma_{1}%
\end{array}\right). \label{Darboux2'a}
\end{equation}%
with the gauge fields $(B_{1},B_{0})$ transforming as
\begin{eqnarray}
B_{1}'= D B_{1} D^{-1}+D_{x}D^{-1},\quad \quad \quad B_{0}'= D
B_{0} D^{-1}+D_{t}D^{-1}.\label{MatrixDarboux3b}
\end{eqnarray}
In either case, it is easy to determine that \eqref{MatrixDarboux3} or \eqref{MatrixDarboux3b} lead to
\begin{eqnarray}
J'_{0}= J_{0}, \quad\quad\quad
J'_{1}=J_{1}-J_{0x}+2\sigma_{1x}=J_{1}-J_{0x}+2\partial^{2}\ln \psi_{1},
\end{eqnarray}
which coincides with \eqref{Darboux3}. The matrix Darboux transformation can also be iterated $N$ times to obtain expressions for the $N$-fold Darboux transformation on the
sets $(\Psi, A_{1}, A_{0})$ and $(\Psi, B_{1}, B_{0})$ which lead to the same results as in the scalar case.

\section{Modified Darboux transformation}

As we have seen in the last section, the conventional Darboux transformation, in the case of the TB hierarchy, generates new solutions starting with known ones (see \eqref{DTfield}-\eqref{DTfieldb}). However, it seems to be quite restricted in the sense that the variable $J_{0}$ does not seem to transform at all under such a transformation. In this section we present a modified Darboux transformation of the linear system (\ref{linear4}) such that the requirement of covariance of the linear
system (\ref{linear4}) allows one to create new solutions of the
TB system (\ref{TB2}) which are more general (namely, it allows for $J_{0}$ to transform).

Let us consider the transformation of the linear system (\ref{linear4}) defined by
\begin{eqnarray}
&& \psi(x,t;\lambda )\longmapsto
\psi'(x,t;\lambda,\lambda_{1})=\psi-\sigma^{-1}_{1}\psi_{x}.\label{Darbouxlike1}
\end{eqnarray}%
The new solution $\psi'(x,t;\lambda,\lambda_{1})$ vanishes at $\lambda=\lambda_{1}$, namely,  $\psi'(x,t;\lambda,\lambda_{1})|_{\lambda=\lambda_{1}}=0$ as in the conventional Darboux transformation in \eqref{Darboux2}. In fact, we note that the modified Darboux transformation \eqref{Darbouxlike1} is related to the conventional Darboux transformation \eqref{Darboux1} through a (space-time dependent) factor $(-\sigma_{1}^{-1})$, namely,
\begin{equation}
\psi'_{\rm modified} = - \sigma_{1}^{-1} \psi'_{\rm conventional}.\label{modified}
\end{equation}
However, this is sufficient to modify the character of the transformations for the potentials.

Requiring the transformed function $\psi'(x,t;\lambda,\lambda_{1})$ to satisfy the linear system (covariance)
\begin{eqnarray}
\psi'_{xx}=\left(J'_{0}+\lambda\right)\psi'_{x}+\left(J'_{0x}-J'_{1}\right)\psi',\quad\quad\quad
\psi'_{t}=\left(J'_{0}-\lambda\right)\psi'_{x}+\left(J'_{1}-J'_{0x}\right)\psi',\label{DTlikelinear2}
\end{eqnarray}
determines the transformations of $J_{0},J_{1}$ under the modified Darboux transformation to be
\begin{eqnarray}
J'_{0}&=&J_{0}-\frac{\sigma_{1x}}{\sigma_{1}},\label{Darbouxlike3a}\\
J'_{1}&=&J_{1}-J_{0x}+\sigma_{1x}. \label{Darbouxlike3b}
\end{eqnarray}
It is interesting to note that, unlike the case of the conventional Darboux transformation (see \eqref{DTfield}-\eqref{DTfieldb}), here both $J_0$ and $J_{1}$
transform under the action of the modified Darboux
transformation (\ref{Darbouxlike1}) and, therefore, it has a richer structure. These new potentials satisfy the evolution equation (which follows from the compatibility of \eqref{DTlikelinear2})
\begin{eqnarray}
J'_{0t}=\left(2J'_{1}+(J'_{0})^{2}-J'_{0x}\right)_{x},\quad\quad
J'_{1t}=\left(2J'_{0}J'_{1}+J'_{1x}\right)_{x},\label{DTlikeTB2}
\end{eqnarray}
which is the TB equation \eqref{TB2} in the new variables. Namely, $(J'_{0},J'_{1})$ can be thought of as new solutions of the TB equation starting from known ones $(J_{0},J_{1})$.

In other words we can say that if the set
$\left(\psi,J_{0},J_{1}\right)$ represents a solution of the linear system
(\ref{linear4}), then the set $\left(\psi[1],J_{0}[1],J_{1}[1]\right)$  given by
\begin{eqnarray}
\psi[1]&\equiv&\psi-\frac{\psi_{1}}{\psi_{1x}}\psi_{x}=\frac{W(\psi,\psi_{1})}{\psi_{1x}},\label{RDarboux5}\\
J_{0}[1]&\equiv& J_{0}-\frac{\sigma_{1x}}{\sigma_{1}}=J_{0}-\partial \ln \left(\frac{\psi_{1x}}{\psi_{1}}\right) ,\label{DTfield1}\\
J_{1}[1]&\equiv&
J_{1}-J_{0x}+\sigma_{1x}=J_{1}-J_{0x}+\partial^{2}\ln \psi_{1},
\label{DTfield2}
\end{eqnarray}%
the one-fold (modified) Darboux transformation under which \eqref{linear4} is covariant. Following the discussion of the last section, the two-fold Darboux transformation of
the set $\left(\psi,J_{0},J_{1}\right)$ can be constructed and is given by
\begin{eqnarray}
\psi[2]&\equiv&\psi[1]-\frac{\psi_{2}[1]}{\left(\psi_{2}[1]\right)_{x}}\left(\psi[1]\right)_{x}
=\frac{W(\psi_{1},\psi_{2},\psi)}{W(\psi_{1x},\psi_{2x})},\label{twoRDarboux1a}\\
J_{0}[2]&\equiv&J_{0}[1]-\partial \ln\left(\frac{\left(\psi_{1}[1]\right)_{x}}{\psi_{1}[1]}\right)=J_{0}-\partial\ln\left(\frac{W(\psi_{1x},\psi_{2x})}{W(\psi_{1},\psi_{2})}\right), \label{twoDTfield1}\\
J_{1}[2]&\equiv&J_{1}[1]-J_{0x}[1]+\partial^{2}\ln \psi_{2}[1]=
J_{1}-2J_{0x}+\partial^{2}\ln W(\psi_{1},\psi_{2}),
\label{twoDTfield2}
\end{eqnarray}%
where $\psi_{i}$ with $i=1,2$ denote solutions of the linear system (\ref{linear4}) corresponding to the eigenvalues $\lambda_{i}$ and the Wronskians are defined in \eqref{wronskian1} and \eqref{wronskian2}.

The (modified) Darboux transformation can be iterated $N$-times to obtain the
$N$-fold Darboux transformation of the set $\left(\psi,J_{0},J_{1}\right)$ which has the form
\begin{eqnarray}
\psi[N]&=&\frac{W(\psi_{1},\psi_{2},\dots,\psi_{N},\psi)}{W(\psi_{1x},\psi_{2x},\dots,\psi_{Nx})},\label{nfoldRDarboux1a}\\
J_{0}[N]&=&J_{0}-\partial\ln\left(\frac{W(\psi_{1x},\psi_{2x},\dots,\psi_{Nx})}{W(\psi_{1},\psi_{2},\dots,\psi_{N})}\right), \label{nfoldDTfield1}\\
J_{1}[N]&=&J_{1}-N J_{0x}+\partial^{2}\ln
W(\psi_{1},\psi_{2},\dots,\psi_{N}), \label{nfoldDTfield2}
\end{eqnarray}%
with the Wronskian defined in \eqref{wronskian3}. From \eqref{NDTfieldb} as well as \eqref{nfoldDTfield2} it follows that in either case the multi-soliton solutions can be expressed in terms of Wronskians.

Following the connection between the scalar and matrix descriptions for the TB hierarchy \cite{daslecture}, we can now define the modified Darboux transformation for the
matrix wavefunction $\Psi$ of the linear system (\ref{zero2}) or \eqref{zero7}. In fact, the generic form of the transformation in this case can be written as (compare with \eqref{MatrixDarboux1a} and \eqref{modified})
\begin{equation}
\Psi'(x,t;\lambda,\lambda_{1})\equiv D(x,t;\lambda,\lambda_{1}
)\Psi(x,t;\lambda )=\left(\mathbbm{1} -M^{-1}\partial\right)\Psi(x,t;\lambda
), \label{ModifiedMatrixDarboux1}
\end{equation}%
where $M$ has the form given in \eqref{Darboux2a} or \eqref{Darboux2'a} depending on the matrix description. The transformation of the (gauge) potentials continues to be given by \eqref{MatrixDarboux3} or \eqref{MatrixDarboux3b} for the covariance of the linear matrix equation and this leads explicitly to (\ref{Darbouxlike3a}) and (\ref{Darbouxlike3b}).

\section{Explicit one soliton/kink solutions}

\begin{figure}[here]
\centering
\includegraphics[width=0.50\textwidth ]{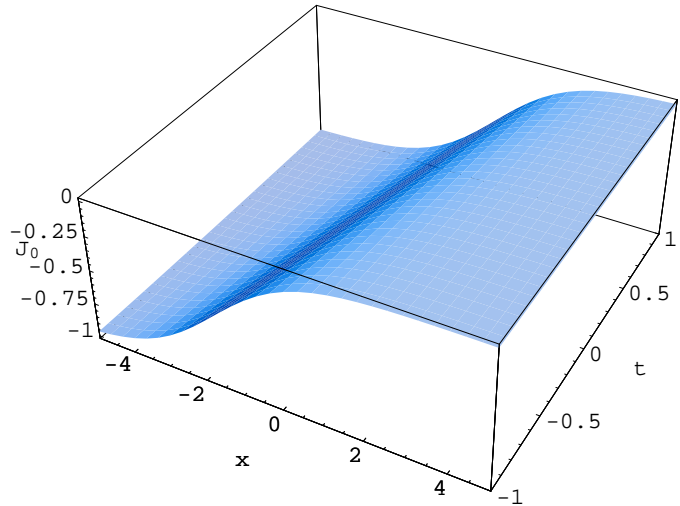}
\caption{One kink solution $J_{0}[1]$ of the TB system given by equation
(\ref{solution2}) with  $\lambda_{1}=1, a=1,b=2$.} \label{fg1}
\end{figure}

\begin{figure}[here]
\centering
\includegraphics[width=0.50\textwidth ]{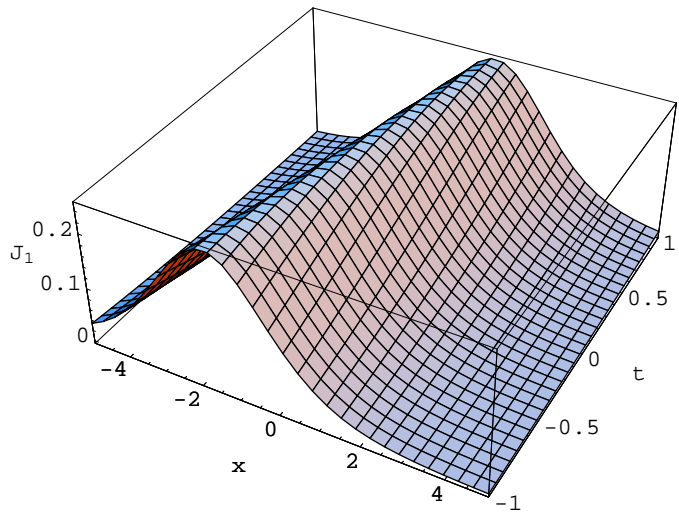}
\caption{ One soliton solution $J_{1}[1]$ of the TB system given by equation
(\ref{solution3}) with  $\lambda_{1}=1, a=1,b=2$.} \label{fg2}
\end{figure}

We can now calculate explicitly the one soliton/kink solutions of the TB system (\ref{TB2}), using the
Darboux transformations. For example, we note that with the seed (trivial) solutions $J_{0}=0$ and
$J_{1}=0$, the linear system has the general solution of the form
\begin{eqnarray}
\psi(x,t;\lambda)&=& a+be^{\lambda(x-\lambda t)},\label{solution1}
\end{eqnarray}
where $a,b$ are constants and $\lambda$ is the spectral parameter. It follows now that
\begin{equation}
\sigma_{1} = \frac{\psi_{1x}}{\psi_{1}} = \frac{\lambda_{1}b e^{\lambda_{1} (x-\lambda_{1}t)}}{a + b e^{\lambda_{1} (x-\lambda_{1}t)}},
\end{equation}
and the modified Darboux transformations (\ref{DTfield1}) and (\ref{DTfield2}) lead to
\begin{eqnarray}
J_{0}[1]&=&-\frac{\lambda_{1} a}{a+be^{\lambda_{1}(x-\lambda_{1} t)}}, \label{solution2}\\
J_{1}[1]&=&\frac{\lambda_{1}^{2}b}{\left(ae^{-\left(\frac{\lambda_{1}
x-\lambda_{1}^2 t}{2}\right)}+be^{\left(\frac{\lambda_{1}
x-\lambda_{1}^2 t}{2}\right)}\right)^{2}}.\label{solution3}
\end{eqnarray}
The solutions of the TB equation in (\ref{solution2}) and
(\ref{solution3}) can be plotted, say for example for $\lambda_{1}=1, a=1, b=2$, and are shown in
Fig. \ref{fg1} and Fig. \ref{fg2} respectively. We see that the solution
(\ref{solution2}) for $J_{0}$ represents a one kink solution (Fig. \ref{fg1}) while the solution (\ref{solution3}) for $J_{1}$ corresponds to a one soliton solution (Fig. \ref{fg2}).

This soliton/kink behavior of the solutions is manifest analytically for $a=1$ and $b=1$ in which case  (\ref{solution2}) and (\ref{solution3}) take the forms
\begin{eqnarray}
J_{0}[1]&=&-\frac{\lambda_{1}}{2}+\frac{\lambda_{1}}{2}\tanh\left(\frac{\lambda_{1}
x-\lambda_{1}^2 t}{2}\right), \label{solution4}\\
J_{1}[1]&=&\frac{\lambda_{1}}{4}{\rm sech}^2\left(\frac{\lambda_{1}
x-\lambda_{1}^2 t}{2}\right).\label{solution5}
\end{eqnarray}
These particular solutions (\ref{solution4}) and (\ref{solution5})
coincide with the solutions of the TB system obtained in
\cite{TBsystem0}-\cite{TBsystem12} through Hirota's method. We
note here that we can also construct soliton/kink solutions of the
TB equation using (\ref{DTfielda}) and (\ref{DTfieldb}). However,
in this case, these solutions would only be for the variable
$J_{1}$.

\section{Concluding Remarks}

In this paper we have studied the Darboux transformations for the
TB hierarchy both in the scalar as well as the matrix descriptions
of the linear equation. While the Darboux transformations have
been extensively studied within the context of AKNS systems based
on $SL(2,R)$, this is the first model where the symmetry group
corresponds to $SL(2,R)\otimes U(1)$. The relation between the
scalar and the matrix descriptions in the present case implies
that the generic form of the Darboux transformation in the matrix
case is different for the TB hierarchy. We show that the
conventional Darboux transformation is quite restricted in this
case in the sense that one of the dynamical variables remains
inert under the transformation. We construct a modified Darboux
transformation which has a richer structure and allows for change
in both the dynamical variables of the theory. We show that in
both the conventional as well as the modified Darboux
transformations, the $N$-fold transformations (multi-soliton
solutions) can be expressed in terms of Wronskians. We have
constructed explicit one soliton/kink solutions for this model
using the modified Darboux transformation. The generalization of
these results to the case of supersymmetric TB hierarchy is
currently under study.

\bigskip

\noindent {\bf Acknowledgement:\\} US would like to thank the theory group at the Department of Physics and Astronomy, University of Rochester for hospitality and the Higher Education
Commission, Pakistan for a research fellowship which made this work possible. He
would also like to thank Mahmood ul Hassan for discussions and valuable
comments. This work was supported in part by US DOE Grant number DE-FG 02-91ER40685.

\end{document}